%% file: guide.tex
\documentclass[letterpaper, 14 pt]{article}

\usepackage{amsfonts}
\usepackage{amsmath}
\usepackage{amssymb}
\usepackage{amsthm}

\usepackage[english]{babel}
\usepackage{graphicx}

\usepackage{hyperref}
\hypersetup{
    colorlinks,
    linktoc=all,
    citecolor=black,
    filecolor=black,
    linkcolor=black,
    urlcolor=blue
}
\makeatletter
\renewcommand{\@seccntformat}[1]{%
  \ifcsname prefix@#1\endcsname
    \csname prefix@#1\endcsname
  \else
    \csname the#1\endcsname\quad
  \fi}
\newcommand\prefix@section{Section \thesection: }
\makeatother
\usepackage{listings}
\usepackage{setspace}
\DeclareFixedFont{\ttb}{T1}{txtt}{bx}{n}{12} 
\DeclareFixedFont{\ttm}{T1}{txtt}{m}{n}{12}  

\usepackage{color}
\definecolor{deepblue}{rgb}{0,0,0.5}
\definecolor{deepred}{rgb}{0.6,0,0}
\definecolor{deepgreen}{rgb}{0,0.5,0}
\definecolor{purple}{rgb}{0.6,0,0.6}
\definecolor{deeporange}{rgb}{0.8,0.5,0}
\usepackage[table]{xcolor}

\newcommand\pythonstyle{\lstset{
language=Python,
basicstyle=\ttm,
otherkeywords={self},             
keywordstyle=\ttb\color{deepblue},
emph={MyClass,__init__,assert},          
emphstyle=\ttb\color{deeporange},    
stringstyle=\color{deepgreen},
commentstyle=\color{purple},
frame=tb,                         
showstringspaces=false            %
}}

\lstnewenvironment{python}[1][]
{
\pythonstyle
\lstset{#1}
}
{}

\newcommand\bashstyle{\lstset{
language=bash,
basicstyle=\ttm,
numbers=none,
otherkeywords={self},             
keywordstyle=\ttb\color{deepblue},
emph={pip,python},          
emphstyle=\ttb\color{deepgreen},    
stringstyle=\color{deepgreen},
commentstyle=\color{purple},
frame=tb,                         
showstringspaces=false            %
}}

\lstnewenvironment{bash}[1][]
{
\bashstyle
\lstset{#1}
}
{}

\begin{document}
\doublespace

\begin{titlepage}
\clearpage
\title{\Large \bf Dynamic Gauss Newton Metropolis Algorithm 
\\ \large The MCMC Jagger}

\author{
		Mehmet Ugurbil }
\date{}
\maketitle

\centering 
\singlespace

\

\

New York University, Courant Institute of Mathematical Sciences

\

251 Mercer Street

New York, NY 10012

01/2016

\

\

\

\

\

A thesis submitted in partial fulfillment 

of the requirements for the degree of

Master of Science 

Department of Mathematics

New York University

01/2016

\

\

\

\

\

\

\_\_\_\_\_\_\_\_\_\_\_\_\_\_\_\_\_\_\_\_\_\_\_\_\_\_\_\_

Jonathan Goodman
					
\thispagestyle{empty}

\end{titlepage}

\tableofcontents
\newpage

\section*{Abstract}

GNM: The MCMC Jagger. A rocking awesome sampler. 

This python package is an affine invariant Markov chain Monte Carlo (MCMC) sampler based on the dynamic Gauss-Newton-Metropolis (GNM) algorithm. The GNM algorithm is specialized in sampling highly non-linear posterior probability distribution functions of the form $e^{-||f(x)||^2/2}$, and the package is an implementation of this algorithm.

On top of the back-off strategy in the original GNM algorithm, there is the dynamic hyper-parameter optimization feature added to the algorithm and included in the package to help increase performance of the back-off and therefore the sampling. Also, there are the Jacobian tester, error bars creator and many more features for the ease of use included in the code. 

The problem is introduced and a guide to installation is given in the introduction. Then how to use the python package is explained. The algorithm is given and finally there are some examples using exponential time series to show the performance of the algorithm and the back-off strategy.

\newpage

\input{intro.tex}

\input{code.tex}

\input{algo.tex}

\input{examples.tex}


\clearpage

\section{References}

\hspace{12pt} [1] B. Zhu. {\it Gauss-Newton-Metropolis with backup strategy}. Courant Institute of Mathematical Sciences, New York University, New York, NY, 2013. 

[2] J. Nocedal and S. Wright. {\it Numerical Optimization}. Springer, New York, 2000.

[3] J.  Goodman  and  J.  Weare. {\it Ensemble   samplers   with   affine   invariance}. CAMCS, Vol. 5 (2010), No. 1, pp. 6580.

[4]  W.  R.  Gilks,  S.  Richardson  and  D.  J.  Spiegelhalter. {\it Markov  Chain  Monte Carlo in practice}. Chapman, 1996.

\end{document}

%% file: intro.tex
\section{Introduction}

\subsection{The Problem}

The GNM algorithm is for sampling generalizations of probability densities of the form $p(x)\propto e^{-||f(x)||^2/2}$. Here $x\in \mathbb{R}^n$ is a ``parameter" vector, $f:\mathbb{R}^n\to\mathbb{R}^m$ is a ``model" function, and $||f(x)||^2=\sum_{k=1}^m f_k^2(x)$. Typically $m>n$. GNM is a powerful algorithm for ill-conditioned problems because it is affine invariant. 

This is related to nonlinear least squares $\min_x ||f(x)||^2$ that arises in statistics. For example, if $m(x)$ predicts the outcome of an experiment with parameter $x$, and $y_k$ are measurements, then the goodness of fit is $\sum_{k=1}^m\frac{(m_k(x)-y_k)^2}{2\sigma_k^2}$. If $f_k(x)=\frac{m_k(x)-y_k}{\sigma_k}$, then this sum has the form $||f(x)||^2/2$.

One generalization to the probability density function $p(x)$ is adding a Gaussian prior probability distribution on the parameters, $\pi:\mathbb{R}^n\to\mathbb{R}$, such that $\pi(x)=\frac{1}{Z}e^{-(x-m)^T H(x-m)/2}$, where $m$ is the mean and $H$ is the precision matrix, the inverse of the covariance. Another generalization is having an indicator function $\chi:\mathbb{R}^n\to \{0,1\}$ to limit the domain of the model function.

The user of the sampler must write Python code to evaluate $\chi(x)$, $f(x)$ and $\nabla f(x)$, the Jacobian. It is optional to add prior information, mean vector $m\in\mathbb{R}^n$ and precision matrix $H\in\mathbb{R}^{n\times n}$ (inverse of the covariance matrix). 

\subsection{Installation}

The {\bf gnm} is a Python package. The Python package \href{http://www.numpy.org}{\textbf{numpy}} is required before the execution of \textbf{gnm}. To use the examples and plot the results, you will need \href{http://matplotlib.org}{\textbf{matplotlib}}. To use the \textit{acor} feature, you will need the package \href{http://www.math.nyu.edu/faculty/goodman/software/acor/}{\textbf{acor}}.

From the default Python packages, you will need \textbf{os}, \textbf{setuptools} (or \textbf{distutils}), \textbf{re}, \textbf{sys}, \textbf{copy}, and \textbf{json}. These packages likely come with your Python installation. 

The easiest way to install \textbf{gnm} would be to use \href{https://pypi.python.org/pypi/pip}{pip}. 

\begin{bash}
$ pip install gnm
\end{bash}

To download manually, use git clone or download as zip from \url{https://github.com/mugurbil/gnm}. 

\begin{bash}
$ git clone https://github.com/mugurbil/gnm.git
\end{bash}

Then you can install manually by going into the \textbf{gnm} directory, and then running \textsf{setup.py}.

\begin{bash}
$ python setup.py install
\end{bash}

To clean the repository after installation, one can run clean with 
\textsf{setup.py}.

\begin{bash}
$ python setup.py clean
\end{bash}

\newpage

%

%% file: code.tex
\section{Code}


\subsection{User Function}
The user needs to create Python code that contains the model information. This Python function, which we will call the user function, should have two inputs, and three outputs. This code will be passed to the sampler.

The first input should be the parameter vector, $x\in\mathbb{R}^n$. The second input should be the arguments that the function takes, which may include the data. 


The first output is the constraint ${\chi(x)}$. It is supposed to tell where the model function is defined. Thus, it would have value $1$ (or True) when the function and its derivative are defined, and $0$ (or False) otherwise. If $D$ is the domain of $f$ and $\nabla f$, then ${\chi(x)}=\left\{
	\begin{array}{ll}
		1  & x\in D \\
		0 &  x\not\in D
	\end{array}
\right.$. The second output is an array of size $m$ of $f$ evaluated at $x$, $f(x)\in\mathbb{R}^m$ (range), that should be convertible to an {\bf np.array}. The third output is an array of size $m$ by $n$ that is the Jacobian matrix, $\nabla f$, evaluated at $x$, $J(x)\in\mathbb{R}^{m\times n}$. This also needs to be convertible to an {\bf np.array}. 

\begin{python}[caption=User defined function outline.]
def model(x, args):
   ...
   return chi(x), f(x), J(x)
\end{python}

\subsection{The Sampler}

The sampler needs the user function ({\bf model}), arguments the user function takes ({\bf args}), and the initial guess of the sampler ({\bf x\_0}). This information is provided to the sampler during initialization. The function will be evaluated at the initial guess to check if it is defined at that point. 

\begin{python}[caption=Sampler initialization.]
jagger = gnm.sampler(x_0, model, args)
\end{python}

The prior information is optional. For a Gaussian prior distribution, mean vector $m$ ({\bf m}) and precision matrix $H$ ({\bf H}) need to be provided. 

\begin{python}[caption=Setting the prior.]
jagger.prior(m, H)
\end{python}

\subsection{Jacobian Test}




The \textbf{Jtest} method is to check whether the derivative information and the model function given in the user function match by employing numerical differentiation. It has two required inputs, six optional inputs, and one output.

The domain, $D_J$, for \textbf{Jtest} should be a rectangle given by two vectors, the top right and the bottom left corners of the rectangle, $\textbf{x}_{\text{max}}$ and $\textbf{x}_{\text{min}}$ respectively. These vectors are the two required inputs. Note that they should have the same dimensions, and all values of $\textbf{x}_{\text{min}}$ should be less than the corresponding values of $\textbf{x}_{\text{max}}$ so that the domain is nonempty. 
$$D_J=\{{\bf x}: \forall i, (\textbf{x}_{\text{min}})_i<{\bf x}_i < (\textbf{x}_{\text{max}})_i\}$$

The method has one output, \textbf{error}, that tells whether the numerical Jacobian, $D f$, converged to the Jacobian supplied by the user function, $\nabla f$. The output is 0 if the convergence occurred. Otherwise, it is the norm of the difference of the numeric and provided Jacobians, $||D f - \nabla f||$, at the point it failed to converge.

\begin{python}[caption=Jtest usage.]
error = jagger.Jtest(x_min,x_max)
assert error == 0 #Is the Jacobian info correct?
\end{python}

\subsection{Advanced Jtest}

The six optional inputs of \textbf{Jtest} and their default values are given below. We then explain how the method works and where these variables come into use. 

\begin{table}[h]
\caption{\textbf{Jtest} optional inputs.}
\centering
\setlength{\extrarowheight}{0.1cm}
\begin{center}
    \begin{tabular}{ | l | l | l | }
    \hline
    SYMBOL & IN CODE & DEFAULT VALUE \\[0.2cm] \hline
    $dx$ &\textbf{dx} & $2\cdot10^{-4}$  \\[0.1cm] \hline
    $N$  & \textbf{N} & 1000  \\[0.1cm] \hline
    $\epsilon_{\text{max}}$ & \textbf{eps\_max} & $1\cdot10^{-4}$  \\[0.1cm] \hline
    $p$ & \textbf{p} & 2 \\[0.1cm] \hline
    $l_{\text{max}}$ & \textbf{l\_max} & 50 \\[0.1cm] \hline
    $r$ & \textbf{r} & 0.5 \\[0.1cm]
    \hline
    \end{tabular} 
\end{center}
\end{table}

\textbf{Jtest} chooses a i.i.d. uniform random variable, $\textbf{x}_k$, in its specified domain. Then it calculates the numerical Jacobian at that point by the symmetric difference quotient. This operation is done for all entries in the Jacobian, for $j=1$ to $n$, and for each $j$, $i=1\rightarrow n$. Note that there is no $i$ in the code since the operations are done in vectors. 

$$ (D^{(l)}\textbf{f}(\textbf{x}_k))_{ij}=\dfrac{f_i(\textbf{x}_k+\delta \textbf{x}_j^{(l)})-f_i(\textbf{x}_k-\delta\textbf{x}_j^{(l)})}{2\delta_j^{(l)}}\approx\dfrac{\partial f_i}{\partial x_j}(\textbf{x}_k)$$

Here $\delta\textbf{x}_j^{(l)} = \delta_j^{(l)}*\textbf{e}_j$, where $\textbf{e}_j$ has 0s everywhere but a 1 at the $j$th entry, and $\delta_j^{(l)}$ is the magnitude of the perturbance in $\textbf{e}_j$ direction at the $l$th stage. The initial perturbance magnitude, $\delta_j^{(0)}$, depends on the domain and $dx$ by the equation $\delta_j^{(0)} = (\textbf{x}_{\text{max}}-\textbf{x}_{\text{min}})_j*dx$. 
 
The numerical Jacobian is compared under $\textbf{p}$ norm to the Jacobian given by the user function evaluated at $\textbf{x}_k$, $\epsilon_k^{(l)} = ||D^{(l)}\textbf{f}(\textbf{x}_k)-\nabla\textbf{f}(\textbf{x}_k)||_p$. If this error, $\epsilon_k^{(l)}$, is smaller than $\epsilon_{\text{max}}$, this point passes the differentiation test. Otherwise, the magnitudes of the perturbations are reduced by a multiplication with $r$: $ \delta_j^{(l)} = \delta_j^{(l-1)}*r=\delta_j^{(0)}*r^{l}$, $\forall j$. 

This is repeated for the same point a maximum amount of $l_{\text{max}}$ times, so that $l \leq l_{\text{max}}$. If the point is still generating an error greater than $\epsilon_{\text{max}}$, the Jacobian is said to be incorrect. This is because the numerical Jacobian is not convergent to the provided Jacobian as we decrease $\delta\textbf{x}$ repeatedly, $\delta\textbf{x}\rightarrow {\bf0}$. In this case, the program returns $\epsilon^{(l_{\text{max}})}_k$, the final error at this point. 

This procedure is repeated for $N$ different points. If all the points pass the differential test, then the method returns 0.

\begin{python}[caption=Advanced {\bf Jtest} usage.]
error = jagger.Jtest(x_min,x_max,N=5,dx=0.001)
\end{python}

\subsection{Back-Off}

There are currently two types of back-off strategies, {\bf static} and {\bf dynamic}. Static way is by fixing the maximum number of back-offs as well as the back-off dilation factor. Dynamic way is by fixing the max number of back-offs but changing the dilation factor based on the current position of the sampler. 

\begin{python}[caption=Setting back-off.]
jagger.static(5, 0.2)
# OR 
jagger.dynamic(3)
\end{python}

%
%
%

\subsection{Sampling}
Once all the information is given to the sampler, all it requires is to call the {\bf sample} method with the number of samples wanted ({\bf n\_samples}).

The sampling process can be divided into sessions. If you run the sample method again, the sampler will remember where it left off and continue the sampling from there. Also, you can divide sampling {\bf n\_samples} into {\bf n\_divs} by setting the option {\it divs}={\bf n\_divs}.  

If used on terminal with {\bf n\_divs}$>1$, to see the progress of sampling, {\it visual} option could be set to {\bf True}. Turning {\it visual} to {\bf True} shows the percentage of the sampling completed.

The sampling can be done in safe mode by setting {\it safe}={\bf True}. This will cause the sampler to save progress at every division, and once the sampling is completed.

\begin{python}[caption=Sampling.]
jagger.sample(n_samples, divs=n_divs, 
              visual=True, safe=True) # sample 
\end{python}

The initial guess might be far away from the region where the chain is supposed to be, causing delay before the chain converges to the desired distribution. After the sampling, these undesired initial samples can be burned (discarded) by calling {\bf burn} and providing the number of samples to be burned ({\bf n\_burned}). 

\

\begin{python}[caption=Burning samples.]
jagger.burn(n_burned) # burn
\end{python}

\subsection{Outputs}

Outputs of the sampler can be accessed after the sampling as properties. These outputs are provided in table~\ref*{tab:outputs}. The main output is the {\bf chain}, which contains the Markov chain of the samples. This is an {\bf np.array} of shape $({\bf n\_samples}, n)$. Thus, it has length equal to the number of samples and each sample is a vector of size $n$ containing the position of that sample.

\begin{python}[caption=Information access.]
chain = jagger.chain
\end{python}

\begin{table}[h]
\caption{Outputs.}
\centering
\label{tab:outputs}
\setlength{\extrarowheight}{0.1cm}
\begin{center}
    \begin{tabular}{ | l | l | }
    \hline
    CODE & EXPLANATION \\[0.2cm] \hline
    \textbf{chain} & The Markov chain  \\[0.1cm] \hline
    \textbf{n\_samples} & Number of elements sampled \\[0.1cm] \hline
    \textbf{n\_accepted} & Number of samples accepted \\[0.1cm] \hline
    \textbf{accept\_rate} & Acceptance rate of the algorithm \\[0.1cm] \hline
    \textbf{call\_count} & Number of function calls \\[0.1cm] \hline
    \textbf{step\_count} & Array of size {\bf max\_steps} indicating \\ & how many samples got accepted  \\ & at each back-off step \\[0.1cm] \hline
    \end{tabular} 
\end{center}

\end{table}

There is a method called {\bf error\_bars} for ease of use. It evenly divides the space given by a rectangle into bins and sums up the number of samples that fall into each bin. Also, it estimates the error bars for each bin. It requires three inputs: an integer specifying the number of bins in each dimension ({\bf n\_dims}), a vector specifying the minimum of each dimension of the rectangle ({\bf d\_min}), and a vector specifying the maximum ({\bf d\_max}). These vectors need to be castable to {\bf np.array} and have dimension $n$. The method produces three outputs: the estimate of the posterior ({\bf p\_x}), the estimate of the error bars ({\bf err}), and the midpoint location of each bin ({\bf x}). All three of these outputs are of form {\bf np.array} and of shape $(n,{\bf n\_dims})$. 

\begin{python}[caption=Error bars.]
x,p_x,err = jagger.error_bars(n_dims,d_min,d_max)
\end{python}

\newpage

%% file: algo.tex
\section{Algorithm}

The Gauss-Newton-Metropolis algorithm and the back-off strategy is taken from Zhu [1]. The dynamic back-off is original.

\subsection{Gauss-Newton-Metropolis}


We cannot sample from the posterior distribution directly as it may be highly non-linear. Instead, we can create an approximation, such as a Gaussian approximation, from which we can sample directly. For this purpose, as in the Gauss-Newton algorithm, we are going to replace the value of the function $f$ by its approximation. That is, if we are sampling $z$ from $x$, we are going to replace $f(z)$ by its approximation around $x$: $f(x)+\nabla f(x)(z-x)$. Hence, we get the proposal distribution: 
$$K(x,z)=\frac{1}{Z}\pi(z)e^{-||f(x)+\nabla f(x)(z-x)||^2/2}$$

What may not be clear is that this approximation is Gaussian and therefore can be sampled directly. Up to a factor of $-\frac{1}{2}$, the exponential part of this proposal is: 
$$ (z-m)^T H(z-m)+||f(x)+\nabla f(x)(z-x)||^2$$

After some algebra, as a function of $z$, this may be reduced to: 
$$ z^T(H+||\nabla f(x)||^2)z - 2z^T(Hm-\nabla f(x)^Tf(x)+||\nabla f(x)||^2x)+O(1)$$

Therefore, we can complete the square to get $(z-\mu)^T P(z-\mu)$ by setting:
$$ P= (H+||\nabla f(x)||^2) \text{ and } \mu=P^{-1}(Hm-\nabla f(x)^Tf(x)+||\nabla f(x)||^2x) $$

Thus, we get that $K(x,z)\sim N(\mu, P) $, which is Gaussian and can be sampled directly. Then we use the Metropolis-Hastings algorithm to achieve our original goal of sampling $p(x)$. We need to have an acceptance probability $A$ to satisfy detailed balance: 
$$p(x)K(x,z)A(x,z)=p(z)K(z,x)A(z,x)$$
$$\text{Therefore, } A(x,z) = \min\left\lbrace 1,\frac{p(z)K(z,x)}{p(x)K(x,z)}\right\rbrace$$
\subsection{Back-Off} 

Algorithms that are not globally convergent may have trouble at points that are away from the optimum. The proposal at these points would not be good approximations of the distribution that is being simulated. This leads to a low acceptance probability, and hence a large autocorrelation time. This situation is undesirable and needs to be overcome. Back-off strategy is a way to overcome this difficulty. 

In Markov chain Monte Carlo algorithms, first a proposal is made, then it is accepted according to an acceptance recipe. If it is rejected, the next element in the chain is the same as the previous element. Thus, the chain remains rigid in place. To overcome this, rather than rejecting a sample that is not accepted, the proposal distribution is dilated and a new proposal is made. This is motivated by step size control methods used in line search.

We need to generalize the detailed balance condition and update the acceptance recipe accordingly to accommodate for the back-off. This can be achieved by the ``very detailed balance" and the new acceptance recipe: 
$$ A(x,z_n|z_1,...,z_{n-1})=\min\left\lbrace 1,\frac{p(z_n)K_1(z_n,z_1)[1-A(z_n,z_1)]\cdots K_n(z_n,x)}{p(x)K_1(x,z_1)[1-A(x,z_1)]\cdots K_n(x,z_n)}\right\rbrace$$

The back-off strategy raises the question: how do we choose the dilation factor? The simple way to choose the back-off dilation factor is by setting it to a constant value.

\subsection{Dynamic Back-Off}
A better way to choose the step size is to note where we want the algorithm to sample most. This would be the area with the highest probability, where $p(x)$ has the highest value. We can approximately achieve this by minimizing $||f(x)||^2$ since $p(x)$ is inversely proportional to it:
$$ p(x) = \chi(x)\frac{1}{Z}\pi(x)e^{-||f(x)||^2/2}\propto e^{-||f(x)||^2/2}$$

We can minimize $||f(x)||^2$ in a search direction by polynomial interpolation [2]. If $(z-x)$ is the search direction, then we want to minimize $\phi(t)=||f(x+t(z-x))||^2$ over $t$. This gives $\phi'(t)=2f(x+t(z-x))\nabla f(x+t(z-x)) \cdot (z-x)$. So we know the four values $\phi(0)$, $\phi(1)$, $\phi'(0)$, and $\phi'(1)$ that can be used to employ cubic interpolation and find the approximate minimum. This is used as our dilation factor.
\newpage

%% file: examples.tex
\section{Examples}

\subsection{Quick Start}
\begin{center}
\begin{figure}[h]
\centering
\caption{Sampling results for the simple well problem.}
\includegraphics[scale=0.4]{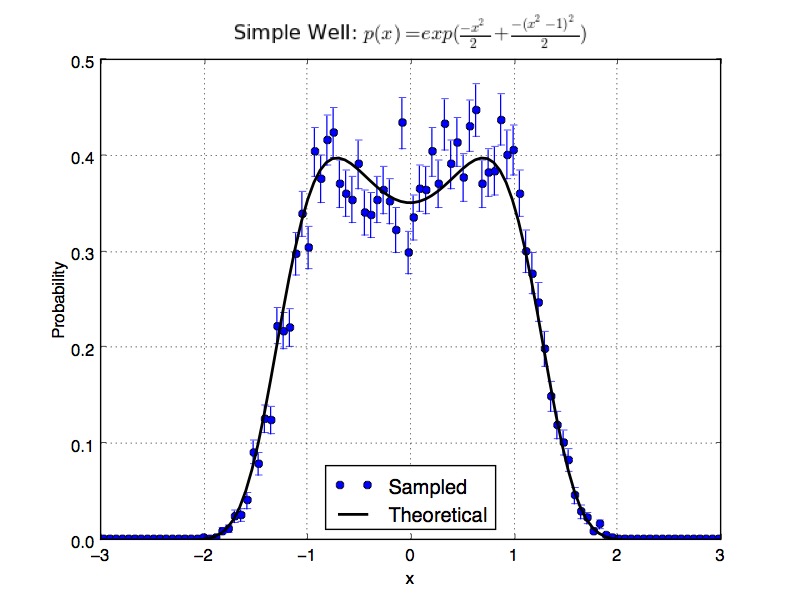}
\label{fig:quick}
\end{figure}
\end{center}

The {\sf quickstart.py} file in {\sf $\sim$/gnm/examples} directory contains a simple example. Running this file should create a plot that looks like Figure~\ref*{fig:quick}.

\begin{bash}
$ python quickstart.py
\end{bash}

In this example, $n=m=1$, $f(x)=\frac{x^2-y}{\sigma}$, $f'(x)=\frac{2x}{\sigma}$. The prior has mean $m=0$ and precision matrix $H=[1]$, giving $\pi(x)=\frac{1}{Z}e^{-x^2/2}$. Therefore, the probability distribution that we want to sample is: 
$$p(x)=\frac{1}{Z}\pi(x)e^{-|f(x)|^2/2}=\frac{1}{Z}e^{-x^2/2-(x^2-y)^2/(2\sigma^2)}$$

\subsection{Well}

The problem in quick-start can be made harder by deepening the well, that is taking $y$ to be bigger. This example is in the {\sf $\sim$/well} folder. The following line will give all the options for the file that you can play around with. 

\begin{bash}
$ python well.py -h
\end{bash}

\subsection{Jtest and Simple 2D}
In this example we first check the usage of Jtest, then sample a simple 2D example. There is a rotator for visiulization so that we can plot any cut of the posterior probability distribution along the plane. 

\subsection{Exponential Time Series}
Suppose we have a process that describes the relationship between time and data by $g(t)=\sum_{i=1}^{n} w_i e^{-\lambda_i t}$. We have measurements $y_k = g(t_k) +\epsilon_k$ for $k=1,...,m$, where the noise is independent and identically distributed and $\epsilon_k\sim N(0,\sigma_k^2)$. Then the parameter is $x=(w_1,...,w_d,\lambda_1,...,\lambda_d)$, with $n=2d$. The model function is given by $f_k(x)=\frac{g_k(t,x)-y_k}{\sigma_k}$ for $k=1,...m$.

\begin{center}
\begin{figure}[h]
\centering
\caption{2D marginal posterior probability function for exponential time series for variables $\lambda_1$ versus $w_1$.}
\includegraphics[scale=0.5]{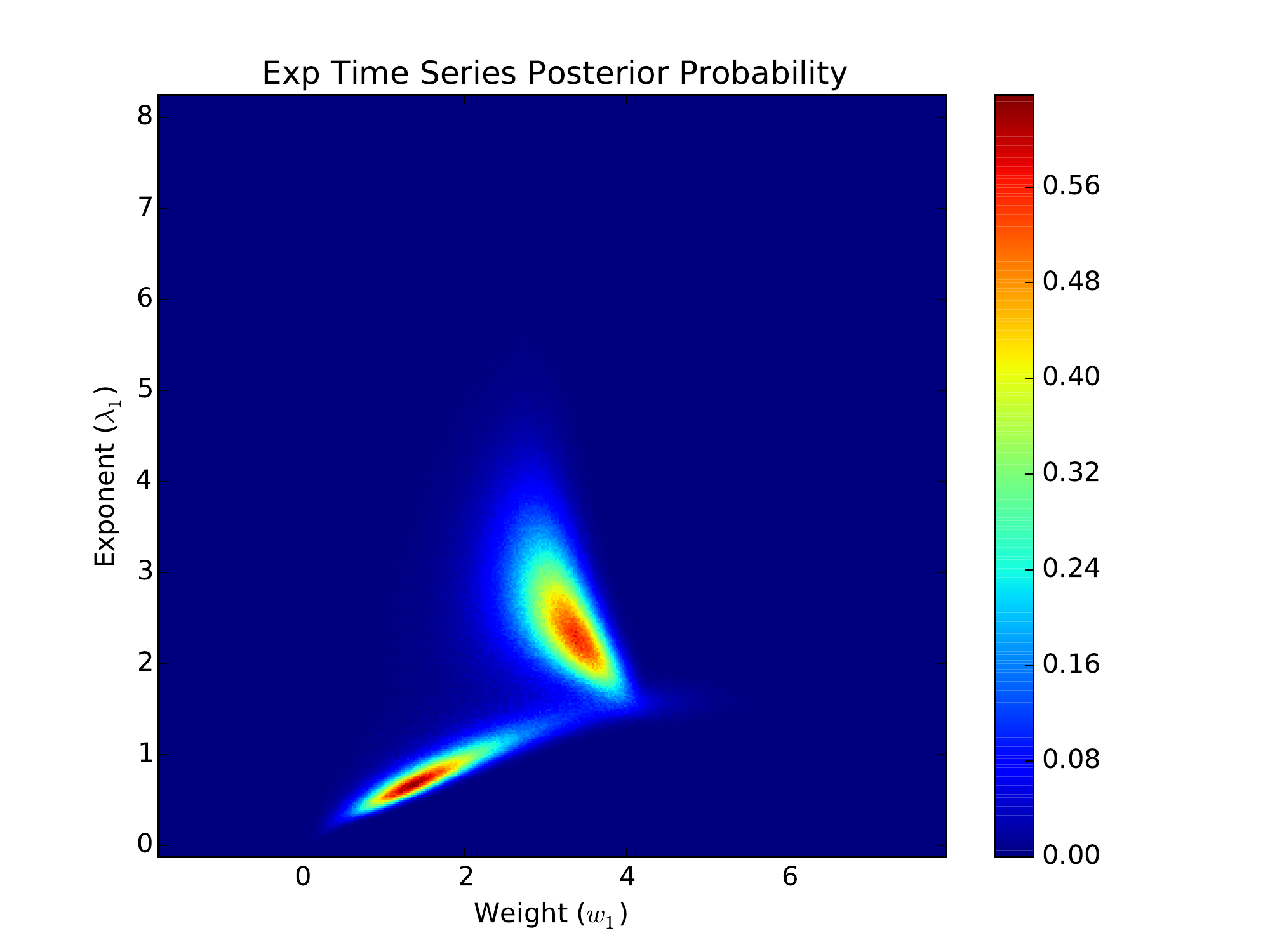}
\end{figure}
\end{center}

Experiment was done for $n=4$ and $m=10$. The sampler was run for $10^5$ samples and first $2000$ was burned in. The mean of the prior was taken to be $[4.,2.,0.5,1.]$ and the precision matrix was chosen as the identity multiplied with $0.5$. The initial guess is the mean of the prior. Data is created by taking the input as $[1.,2.5,0.5,3.1]$ and adding random normal noise with standard deviation $0.1$. 

A 2D marginal histogram of the probability distribution sampled, $\lambda_1$ versus $w_1$, is shown in Figure 2. Figure 3 presents a 1D marginal histogram of the probability distribution sampled, for $\lambda_1$. This figure also displays quadrature (theoretical) curve versus the sampled probability function where we can observe the correctness of the algorithm. 

\begin{center}
\begin{figure}[h]
\centering
\caption{1D marginal posterior probability function for exponential time series for the variable $\lambda_1$.}
\includegraphics[scale=0.5]{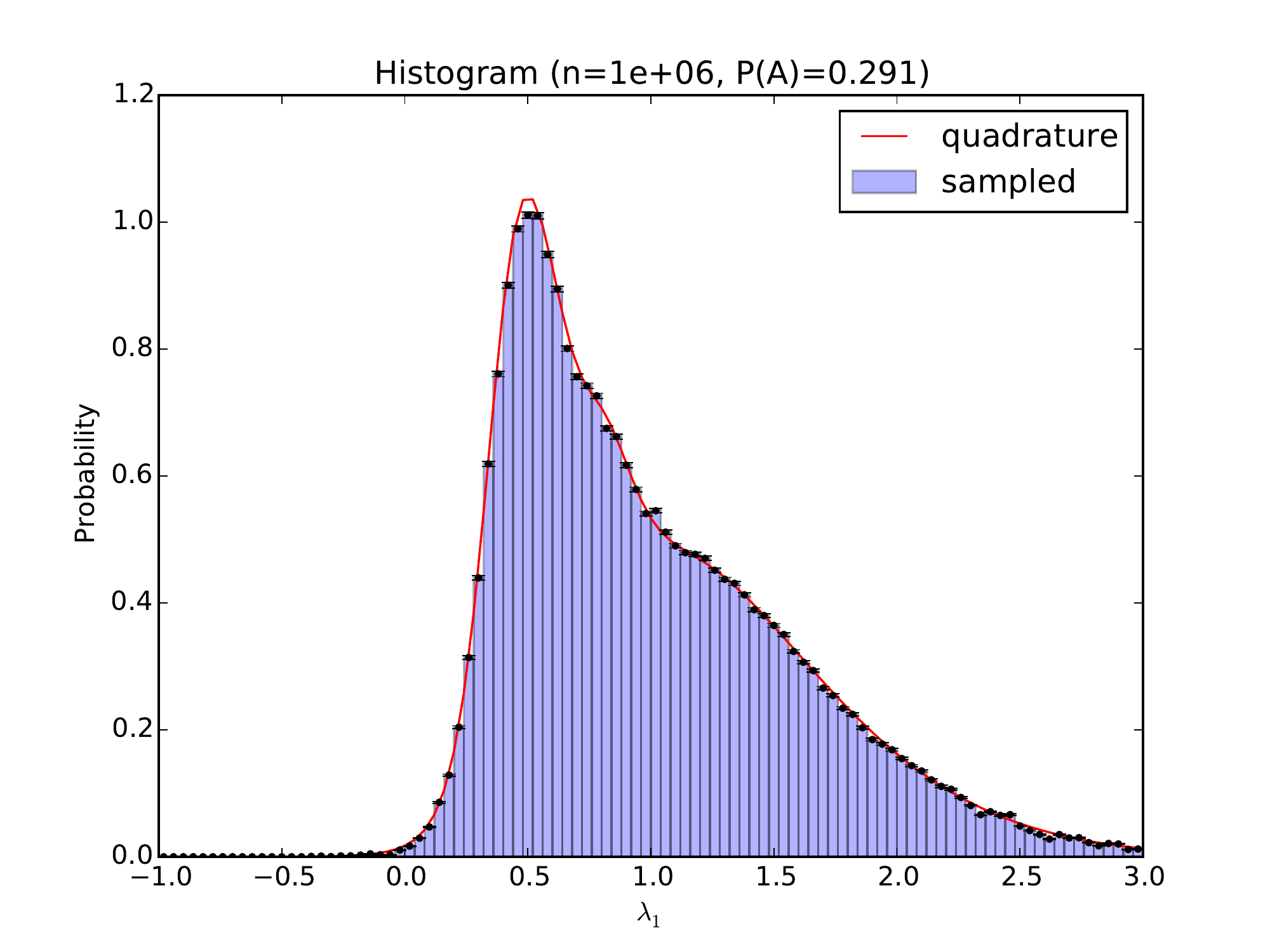}
\end{figure}
\end{center}

\begin{table}[h]
\centering
\caption{Comparison of sampling strategies for sampling exponential time series with different back-off strategies.}
\setlength{\extrarowheight}{0.1cm}
\begin{center}
    \begin{tabular}{ | l| l | l | l | l | l | }
    \hline
    \bf Max & \bf Dilation & \bf Acceptance & \bf Acor & \bf Effective & \bf Number of
    \\[0.1cm] \bf Steps & \bf Factor & \bf Rate & \bf Time & \bf Size & \bf Function Calls
    \\[0.1cm] \hline
     0 & - & 0.273 & 2880 & 3470 & 1.00 * $10^7$\\[0.1cm] 
            \hline
         1 & Dynamic  & 0.653 & 1720 & 5810 &  1.73 * $10^7$\\[0.1cm] 
       \hline
        1 & 0.1  & 0.603 & 1390 & 7180 &  1.73 * $10^7$\\[0.1cm] 
        \hline
        1 & 0.5  & 0.411 & 1760 & 5680 &  1.73 * $10^7$\\[0.1cm] 
       \hline
       2  & 0.1  & 0.812 & 1510 & 6610 & 2.12 * $10^7$ \\[0.1cm] 
        \hline
    
    \end{tabular} 
\end{center}
\end{table}

In this experiment we see significant gains using back-off both in the acceptance rate and in the autocorrelation time after analyzing Table 4, however this conclusion does not hold for the dynamic back-off. Since lower autocorrelation time implies higher effective sample size, we have for instance $8850$ effective samples with 1 back-off step versus $5952$ effective samples for no back-off steps. These translate to $\frac{8850}{1.75*10^5}=0.0506$ versus $\frac{5952}{1.12*10^5}=0.0531$ effective samples per function call implying that the efficiency increases with back-off. This number drops down significantly together with efficiency, to $\frac{9434}{2.66*10^5}=0.0355$, for 3 back-off steps. This suggests that using too many back-offs leads to too many function calls to get the same amount of effective samples, making it inefficient. 

\begin{center}
\begin{figure}[h]
\centering
\caption{Covariance of the first component of the markov chain.}
\includegraphics[scale=0.5]{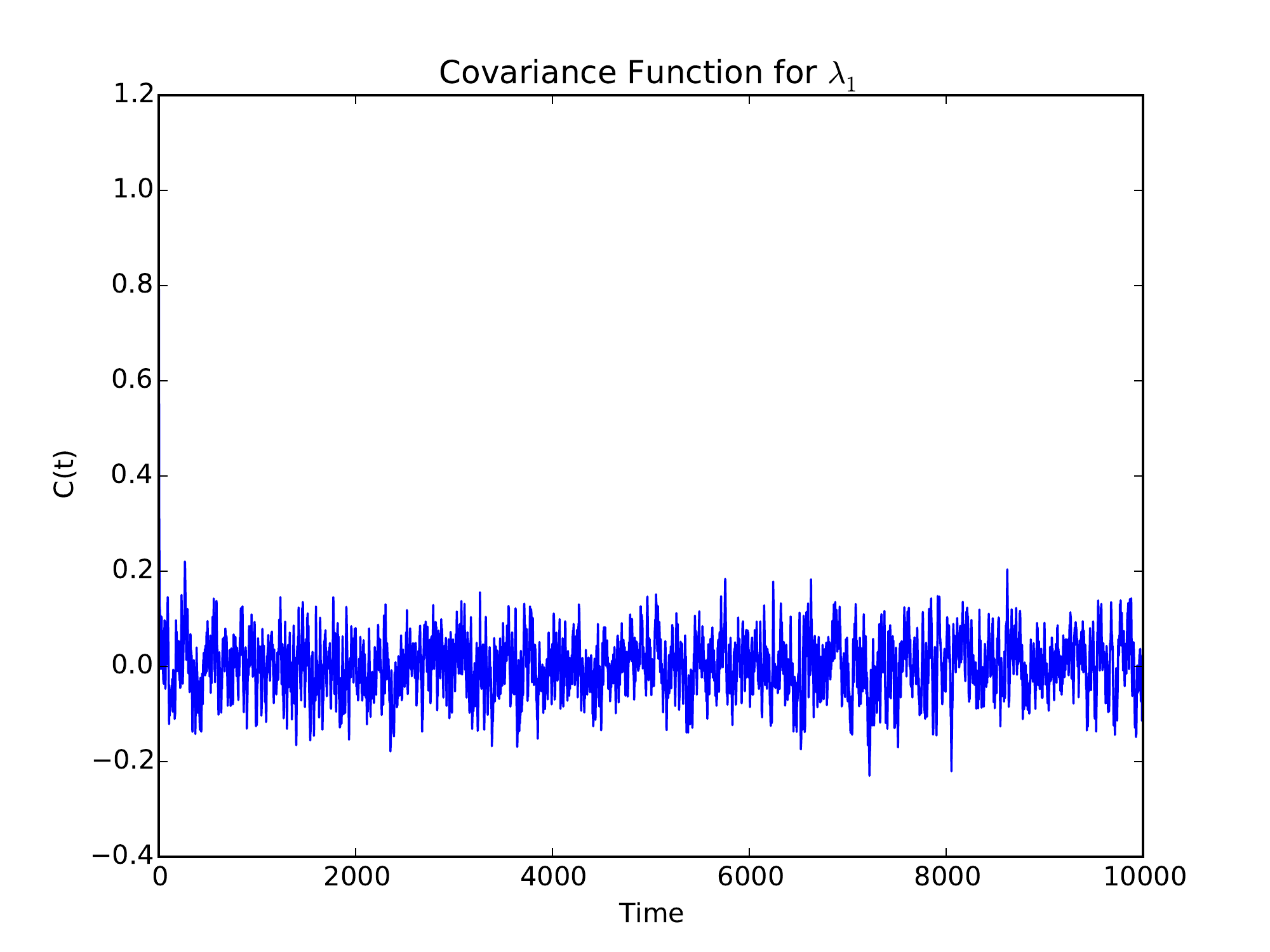}
\end{figure}
\end{center}

In Figure 4, we can observe that the covariance of the chain is stable implying that the chain is stationary. We can see in Figure 5 the percentage of samples accepted at each step (-1 is reject). This example shows that using too many back-off steps does not help significantly, however adding the first back-off step increases the acceptance rate of the algorithm significantly, around 30\%. 

\begin{center}
\begin{figure}[h]
\centering
\caption{Percentage of samples accepted at each back-off step.}
\includegraphics[scale=0.5]{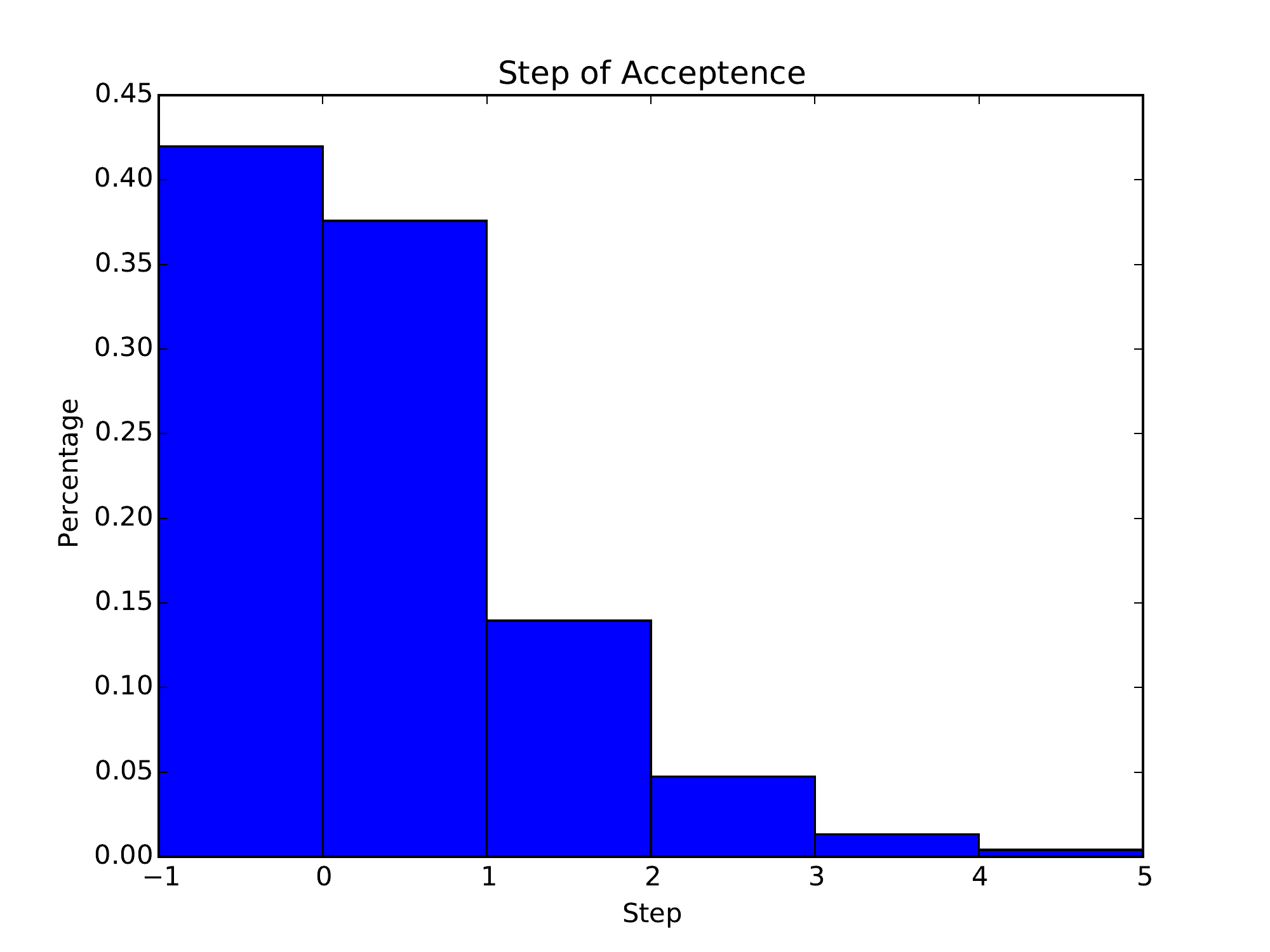}
\end{figure}
\end{center}

\newpage